\documentclass[prl,floatfix,showpacs,superscriptaddress,longbibliography,twocolumn]{revtex4-2}

\usepackage[english]{babel}
\usepackage{amsmath,amssymb,amsfonts}
\usepackage{epsfig}
\usepackage{graphicx}
 \usepackage{relsize}
\usepackage{outlines}
\usepackage{tipa}
\usepackage{braket}
\usepackage[dvipsnames]{xcolor}

\usepackage[colorlinks=true,linkcolor=blue,citecolor=red]{hyperref}
\usepackage[normalem]{ulem}

\newcommand{\figu}[1]
{Fig.~\ref{#1}}

\begin{document}

\title{Fourth-Order Exceptional Points in Correlated Quantum Many-Body Systems}
\author{L.~Crippa}
\affiliation{Institut f\"ur Theoretische Physik und Astrophysik and W\"urzburg-Dresden Cluster of Excellence ct.qmat, Universit\"at W\"urzburg, 97074 W\"urzburg, Germany}

\author{J.~C.~Budich}
\affiliation{Institute of Theoretical Physics, Technische Universit\"at Dresden and and W\"urzburg-Dresden Cluster of Excellence ct.qmat, 01062 Dresden, Germany}

\author{G.~Sangiovanni}
\affiliation{Institut f\"ur Theoretische Physik und Astrophysik and W\"urzburg-Dresden Cluster of Excellence ct.qmat, Universit\"at W\"urzburg, 97074 W\"urzburg, Germany}

\begin{abstract}
Non-Hermtian (NH) Hamiltonians effectively describing the physics of dissipative systems have become an important tool with applications ranging from classical meta-materials to quantum many-body systems. Exceptional points, the NH counterpart of spectral degeneracies, are among the paramount phenomena unique to the NH realm. While realizations of second-order exceptional points have been reported in a variety of microscopic models, higher-order ones have largely remained elusive in the many-body context, as they in general require fine tuning in high-dimensional parameter spaces. Here, we propose a microscopic model of correlated fermions in three spatial dimensions and demonstrate the occurrence of interaction-induced fourth-order exceptional points that are protected by chiral symmetry. We demonstrate their stability against symmetry breaking perturbations and investigate their characteristic analytical and topological properties. 
\end{abstract}
\maketitle

Exploring new physics relating to the ubiquitous occurrence of non-Hermitian (NH) matrices in dissipative settings has become a broad frontier of current research \cite{ashida2020}, prominently including the recent discovery of topological phenomena unique to the NH realm ~\cite{Rudner2009, Zeuner2015, Lee2016, Zhou2018, Lieu2018, Longhi2018, Imhof2018, gong2018, Kunst2018, Yao2018, Yao2018b, Xiong2018, Kawabata2019, Yuce2020, Xiao2020, Weidemann2020, Budich2020, Bergholtz2021}. In general, an effective NH description may not only model open system scenarios~\cite{fukui1998,Menke2017,ashida2020}, but also account for the dissipative environment seen by the individual constituents of a closed system due to disorder ~\cite{Zyuzin2018,michal2019,michen2021} or many-body interactions~\cite{kozii2017,Yoshida2018,Kimura2019,yoshida2020b, Nagai2020,Michishita2020,rausch2021,Lehmann2021}.

Exceptional points, i.e. spectral degeneracies at which the effective NH Hamiltonian becomes non-diagonalizable~\cite{kato1966,berry1994,Heiss2012}, are certainly among the paramount characteristics of NH systems due to their intriguing topological and non-analytical properties~\cite{Heiss2001,Heiss2012,Heiss2016,kozii2017, Carlstrom2018, shen2018a,Yang2021}. Generally speaking, to obtain $n$-th order exceptional points (EP$^n$), where an $n$-fold degenerate eigenvalue corresponds to a single eigenstate, $2n-2$ parameters must be tuned~\cite{holler2020}. Thus, even though higher-order exceptional points have been observed~\cite{melay2018,dey2020,wang2019}, in systems with up to three spatial dimensions, only EP$^2$ occur intrinsically, i.e. without adjusting external parameters. However, various symmetries including PT-symmetry and chiral symmetry have been found to cut the number of parameters to be tweaked 
in half~\cite{budich2019,yoshida2019,delplace2021,mandal2021,bergholtz2021b}, thus suggesting the possibility of naturally observing symmetry-protected exceptional points up to fourth order.

\begin{figure}[ht]
  \includegraphics[width=\linewidth]{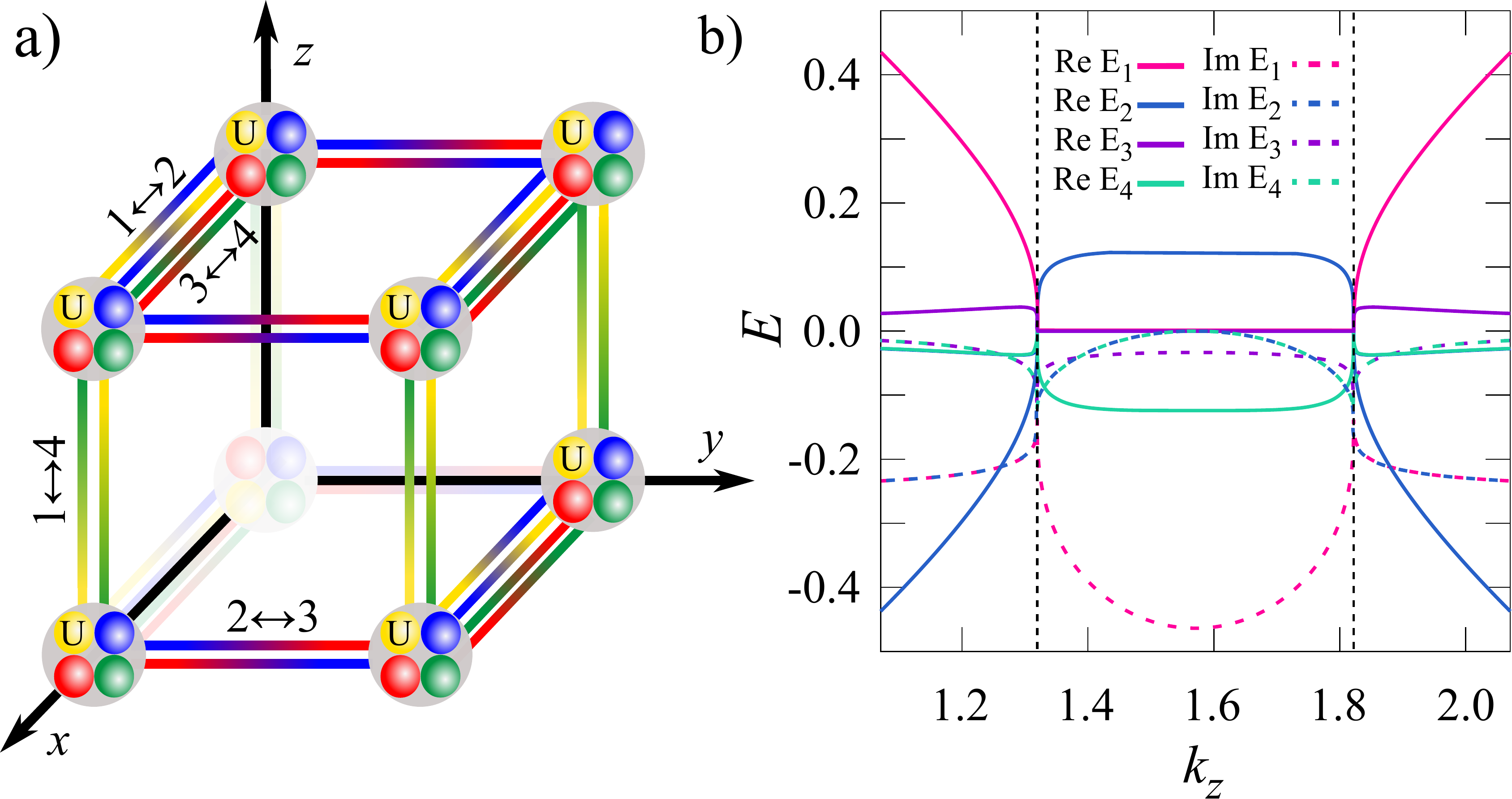}
    \caption{(color online) \textbf{a}) Sketch of the three-dimensional cubic lattice model. Each site hosts four orbitals (differently colored spheres). On one of them (shown in yellow), a local Hubbard repulsion $U$ is active. The hoppings (colored lines) connect the four orbitals along the three directions. \textbf{b}) Band dispersion of the effective NH Hamiltonian as a function of $k_{z}$ when the other coordinates are fixed to those of the EP$^4$. Real and imaginary parts of the eigenvalues $E_{1,2,3,4}$ are degenerate at the $z$-coordinate of the EP$^4$.
  }
  \label{fig1}
\end{figure}

Here, we reveal the occurrence of EP$^4$ induced by electronic correlations in quantum materials that preserve chiral symmetry~\cite{chiu2016, Kawabata2019, yoshida2020b}.
To this end, we propose and analyze a microscopic tight-binding model of a crystalline solid with four orbitals per unit cell and an on-site electron-electron interaction that affects predominantly one of the orbitals (see Fig.~\ref{fig1}a for an illustration). 
We obtain an unbiased solution by means of a non-perturbative many-body approach, which is able to account for the \emph{temperature-dependent} electron-electron scattering of the fermionic quasi-particle excitations. The associated lifetime affects the single-particle levels close to the Fermi energy beyond a standard Boltzmann broadening, leading to the linearized effective NH Hamiltonian
\begin{equation}
\mathlarger{H(\mathbf{k})}\mathlarger{\mathlarger{\mathlarger{=}}}\begin{bmatrix}
i\gamma & {k_x}& 0 & {k_{z}} \\
{k_x} &0&{k_{y}}&  0 \\
0 &  {k_{y}} &0& \text{-}{k_x}\\
{k_{z}}& 0 &\text{-}{k_x} &0\\
\end{bmatrix}\overset{@\text{EP$^4$}}{\mathlarger{\mathlarger{\mathlarger{\simeq}}}}
\begin{bmatrix}
i\frac{\gamma}{4} & 1& 0 &  0\\
0 &i\frac{\gamma}{4} &1&  0 \\
0 & 0&i\frac{\gamma}{4} &1 \\
0 &0 &0&i\frac{\gamma}{4} \\
\end{bmatrix},
\label{nonhermitian_h}  
\end{equation}
where $k$ denotes lattice momentum and $\gamma$ is the anti-Hermitian coupling strength resulting from the interaction-induced self-energy.

The equivalence in (\ref{nonhermitian_h}) refers to a similarity transformation by which $H(\mathbf{k})$ may be brought into a Jordan normal form that makes the EP$^4$, occurring at a finite number of momentum space points, manifest. By contrast to linear level crossings familiar from the Hermitian realm, we show that the quasi-particle dispersion emanating from the observed EP$^4$ indeed exhibits a characteristic fourth root behavior (cf.~Fig.~\ref{fig1}b).  Further, we confirm the robustness of the interaction-induced EP$^4$ of our three-dimensional lattice model against arbitrary chiral-symmetric perturbations. 

\noindent
\textit{Symmetries and stability of exceptional points}--
Degeneracies of eigenvalues in the spectrum of a Hamiltonian are typically achieved by means of fine-tuning of parameters, including both external ones and intrinsic ones such as lattice momentum. The number of constraints necessary to achieve isolated degeneracies is referred to as the \textit{codimension} of the problem. 
Though at first sight it may appear counter-intuitive, for a non-Hermitian system one generally needs less constraints than for its Hermitian counterpart. For example, Weyl points are found in three spatial dimensions (3D) for Hermitian systems~\cite{Murakami2007,turner2013} but EP$^2$ are stable already in 2D in the non-Hermitian case ~\cite{berry2004,yoshida2020b}. 

From a mathematical standpoint, the exceptional points are identified with the degenerate roots of the characteristic polynomial of the Hamiltonian matrix, $P(E)=\mathrm{det}(H-E)$. In particular, for EP$^n$ to be present at a certain position in the parameter space, $P(E)$ and its first $n-1$ derivatives have to simultaneously vanish at that point. These conditions translates into $2n-2$ constraints, already mentioned above, corresponding to the number of complex-valued coefficients~\cite{holler2020}. Hence, EP$^2$ are generally stable in 2D, EP$^3$ in 4D, EP$^4$ in 6D, and so on. The situation is however greatly simplified if  anti-unitary symmetries that are local in momentum space are present. These symmetries relate the Hamiltonian with its complex or Hermitian conjugate, and hence force the coefficients of the characteristic polynomial to be either real or purely imaginary~\cite{delplace2021}. Therefore, the codimension is reduced to $n-1$, which is the reason why in the presence of such symmetries, an EP$^4$ can exist in 3D. 

In our case, the non-Hermitian Hamiltonian of Eq.\eqref{nonhermitian_h} is three-dimensional in the $k$-space and satisfies the chiral symmetry $U_\text{ch}\mathcal{H}(\mathbf{k})U_\text{ch}^{-1}=-\mathcal{H}^{\dagger}(\mathbf{k})$ with the 4$\times$4 unitary matrix $U_\text{ch}=\Gamma_5=\rm{diag}(+1,-1,+1,-1)$. 
Since both the dimensionality of the parameter space and the codimension are equal to 3, the system can support, at most, a finite set of EP$^4$. 
Indeed, as already mentioned, the Hamiltonian becomes non-diagonalizable at $k^{EP}$=$[\pm \text{\textbabygamma}/2\sqrt{2},0,\pm  \text{\textbabygamma}/\sqrt{2}]$, $[\mp \text{\textbabygamma}/4,-\text{\textbabygamma}/4,-3\text{\textbabygamma}/4]$ and $[\mp \text{\textbabygamma}/4,\text{\textbabygamma}/4,-3\text{\textbabygamma}/4]$, where $\text{\textbabygamma}=\gamma/\sqrt{2}$. At these points, all the eigenvalues of the Hamiltonian are degenerate, as it can also be seen in the right panel of \figu{fig1}. Further, a small but finite generic perturbation that respects the chiral symmetry requirement with $\Gamma_{5}$, does not affect the presence of the EP$^4$ but only their positions, simply causing a shift in the $k$-space.

\noindent\textit{Tight-binding realization}-- The non-Hermitian Hamiltonian \eqref{nonhermitian_h} can be seen as a sum of a real symmetric matrix, which is completely off-diagonal, and a diagonal imaginary one. This, along with the fact that the model is three-dimensional, suggests the possibility of finding EP$^4$ protected by the chiral symmetry, in the realm of strongly correlated electron systems. We propose a realization based on the Hermitian non-interacting Hamiltonian of a suitable lattice model, augmented by electron-electron interactions. The dimensionality of the parameter space coincides with that of the lattice, the $k$ variables being the components of the crystal momentum.

Our interacting microscopic Hamiltonian in real space reads
\begin{equation}
\hat{H}\mathsmaller{=}\sum_{\substack{\langle i,j \rangle\\ \alpha, \beta\\\sigma}}t_{\substack{{i,j}\\{\alpha,\beta}\\{\sigma}}}c^{\dagger}_{i\alpha\sigma}c_{j\beta\sigma} \mathsmaller{-}\sum_{i\alpha\sigma}\mu_{\alpha}\hat{n}_{i\alpha\sigma}\mathsmaller{+}\sum_{i\alpha}U_{\alpha}\hat{n}_{i\alpha\uparrow}\hat{n}_{i\alpha\downarrow},
 \label{tb-hamiltonian}   
\end{equation}
where neighboring sites are represented by $\langle i,j \rangle$, the indices $\alpha,\beta=1,2,3,4$ denote the orbitals on each site and $\sigma=\uparrow,\downarrow$ refers to spin. 
Our specific hopping structure is schematically shown in \figu{fig1}a: nearest-neighbor orbital-diagonal hopping is absent; each orbital supports hopping in two spatial directions, $1$ and $4$ along $x$ and $z$, $2$ and $3$ along $x$ and $y$; $t$ is assumed to be uniform in modulus with opposite sign for orbitals $1,2$ and $3,4$ along the $x$ direction. Different spins remain completely decoupled, such that  the single-particle part for each spin block -- apart from an orbital dependent shift of the chemical potential term -- can be written in momentum space as:
\begin{equation}
H_{\sigma}(\mathbf{k})=
\begin{bmatrix}
0 & -\epsilon\cos k_{x} & 0      & -\epsilon\cos k_{z}\\
-\epsilon\cos k_{x} & 0 & -\epsilon\cos k_{y}   & 0   \\
0 & -\epsilon\cos k_{y}   & 0     & \epsilon\cos k_{x}\\
-\epsilon\cos k_{z} & 0     & \epsilon\cos k_{x}   & 0\\
\end{bmatrix}
\label{h-k-space}  
\end{equation}
where $\epsilon=2|t|$ is chosen henceforth as our unit of energy.
At half-filling this model features gap-closing points at the Fermi level, denoted in the following by $\omega=0$. Linearized around the $k$-points $[\pm \pi/2,\pm \pi/2, \pm \pi/2]$ one obtains the Hermitian part of \eqref{nonhermitian_h}. Since the two spin blocks are identical and decoupled, we will from now on only consider one of the two and omit the $\sigma$ index.
Via the third term in Eq.~\eqref{tb-hamiltonian} we introduce electronic interactions into the picture.
As is apparent from its form, we consider the simplest realization of a local Hubbard repulsion, active only when two electrons with opposite spins occupy the same orbital. 

Assuming thermal equilibrium, the Hermitian many-body Hamiltonian (\ref{tb-hamiltonian}) fully characterizes our system as a whole.
Hence, one may wonder where the non-Hermitian description comes into play. 
In this respect, it is important to note that, even though the two-body term is invariant with respect to lattice translations, the crystalline momenta of the individual electrons are not conserved at each scattering event. They are thus entangled because of the presence of the electron-electron interaction.
This is formally accounted for in the context of Landau-Fermi liquid theory~\cite{Abrikosov1959}, which describes low-energy \textit{quasiparticle} excitations as appropriately renormalizated fermions with a finite lifetime.  
The frequency-dependent complex self-energy $\Sigma$ contains all the information on the many-body interaction processes, recast into a single-particle form. In particular, from its imaginary part  at zero frequency, we can extract the finite scattering time for electrons.
Since this effective description holds only close to the Fermi level, one can view the full many-body system as a collection of low-energy quasiparticles coupled to a incoherent background. Interpreting the latter as a reservoir representing the remaining correlation effects establishes the parallel with the physics of coupled open quantum systems.

A convenient way of inspecting the topology of the resulting low-energy many-body problem is to introduce the frequency-dependent \textit{effective} Hamiltonian~\cite{WangZhang2012,kozii2017} $H_{\mathrm{eff}}(\mathbf{k},\omega)=H(\mathbf{k})+\Sigma(\mathbf{k},\omega)$. This fully determines the one-particle Green's function of the system and can be used in a formally similar way to what is done in standard topological band theory~\cite{shen2018a}. 
The quasiparticle lifetime related to $\Sigma(\mathbf{k},\omega)$ has a power-law dependence on the temperature (see, e.g., ~\cite{coleman2015,wagner2021}) rendering the finite-$T$ effective Hamiltonian non-Hermitian. Hence, the combined effect of finite-temperature physics and electronic interaction makes it possible for a system at equilibrium to feature non-Hermitian topological effects such as the appearance of exceptional points~\cite{Michishita2020b}.
In the following, we will specifically consider the effective Hamiltonian at  $\omega=0$, namely $H(\mathbf{k})+\Sigma(\mathbf{k},0)$.

None of the terms in Eq.~\eqref{tb-hamiltonian} breaks chiral symmetry at the operator level \cite{chiu2016}. Therefore, in the absence of spontaneous symmetry breaking, a generic choice of different $U_{\alpha}$ parameters gives rise to a self-energy that still satisfies the chiral symmetry requirement with respect to $\Gamma_{5}$. In general, however, its structure will be $\mathbf{k}$-dependent and off-diagonal in the orbital indices: $\Sigma_{\alpha \beta}(\mathbf{k},\omega)$.
A simplification, still retaining all the desired features, can be obtained by setting $U_{1}$ to a finite value and $U_{2,3,4}=0$.
In this case the self-energy takes indeed the form $\mathrm{diag}(i\gamma,0,0,0)$ and the linearized effective Hamiltonian around $[\pm \pi/2,\pm \pi/2, \pm \pi/2]$ coincides with \eqref{nonhermitian_h}, thereby supporting eight EP$^4$ around each of these $k$-points.
The absence of the real part, bound to vanish at $\omega=0$ irrespective of $k$, is a consequence of the chiral symmetry and can also be proven directly by observing that $\mathrm{Re}\Sigma_{11}$ is odd by construction~\cite{Rohringer2012,Rohringer2013,Note1}.
Further, we are legitimatized to neglect the residual $\mathbf{k}$-dependence in $\gamma$, i.e. the imaginary part of $\Sigma_{11}$. 
Indeed, even if present, this would just result in a more involved algebraic condition for the Hamiltonian to be non-diagonalizable. Thus, the $\mathbf{k}$-dependence would only affect the position of the EP$^4$ without destroying them, provided that it is weak enough. This is the case at high temperature ~\cite{rohringer2016,philipp2017t,gull2011}, where moreover $\mathrm{Im}\Sigma_{11}$ is always non-vanishing.
It is also important to notice that the absence of extended van Hove singularities in the noninteracting density of states~\cite{Note1} guarantees that the system has a weak tendency towards Fermi surface instabilities, which may instead potentially mask the many-body physics of the EP$^4$ presented here.
\begin{figure}[ht]
  \includegraphics[width=\linewidth]{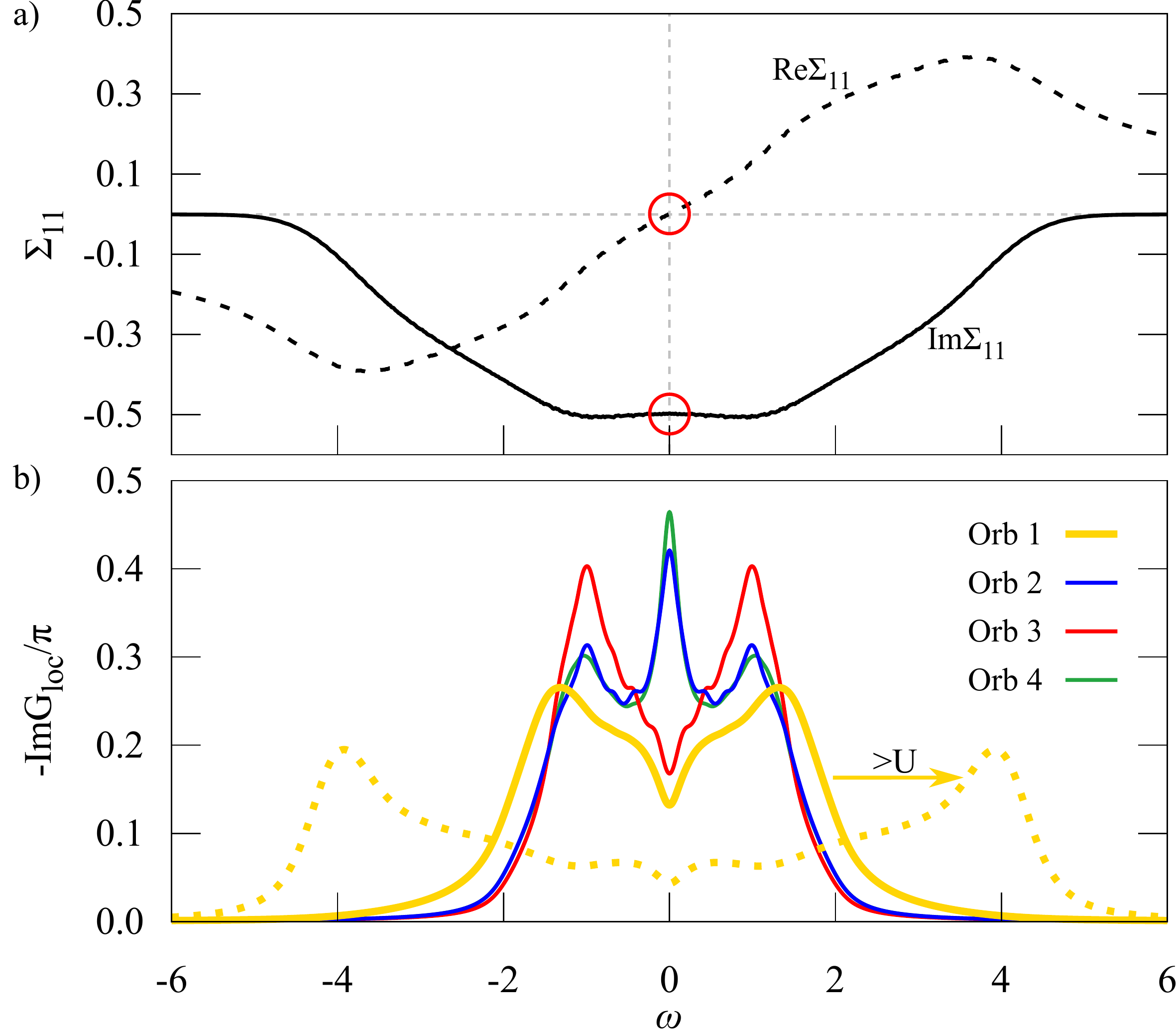}
  \caption{(color online) \textbf{a}: Real and imaginary part of the local self-energy for the correlated orbital on the real frequency axis $\omega$ \textbf{b}: Orbital-resolved local spectral function obtained from Im$G_{\alpha\alpha}(\omega)$ at $U=2$. Upon increasing $U$, the spectral weight of the correlated orbital gets progressively transferred to the high-energy Hubbard bands (see for reference the dashed yellow line at $U=6$).
  }
  \label{fig2}
\end{figure}

Together, all these observations strongly speak in favor of dynamical mean field theory (DMFT) as the ideal tool for solving our many-body lattice problem.
DMFT is an extension of the usual mean-field approach~\cite{GeorgesKotliar1996,MetznerVollhartd1989,MullerHartmann1989} which, by introducing a frequency dependence in the otherwise constant Weiss field, takes all local quantum many-body fluctuations into account. As such, this method is particularly well-suited to describe nonperturbative regimes of interacting fermionic systems in more than 2 spatial dimensions and at finite temperatures.
DMFT goes well beyond Hartree-Fock, in that it is able to describe lifetime effects and temporal, albeit local, many-body fluctuations.
In our specific case, the absence of $k$-dependence in the DMFT self-energy does not however affects any of the non-Hermitian topological properties. This means that for what concerns this specific aspect, DMFT yields a solution of our many-body problem which is \emph{de facto} equivalent to the exact one.

In \figu{fig2} we show the DMFT $\Sigma_{11}(\omega)$ (upper panel) and the local spectral function of each orbital. The spectral features for orbital 1 are strongly sensitive to the value of $U_1$, while the feedback onto the Green's function of the other flavors is less pronounced~\cite{Note1}.
An increase in the value of $U_1$ leads to a progressive spectral weight transfer to high frequencies of the yellow spectrum. This can eventually reach a Mott insulating phase at strong-coupling, but this regime is of no interest to us in this context. 
Therefore, in the following we fix $U_{1}$ to a relatively small value compared to the bandwidth ($\approx 2$) and work at $T\approx1$. Given this choice of parameters, the iterated perturbation theory (IPT)~\cite{georges1992,Kajueter1996,Rozenberg1992} solver for DMFT is reasonably accurate and conveniently allows for a direct access to quantities on the real-frequency axis. As anticipated above, Im$\Sigma_{11}(\omega=0)$ is finite at $T\neq 0$, while the real part vanishes (red circles). 

\begin{figure}[ht]
  \includegraphics[width=\linewidth]{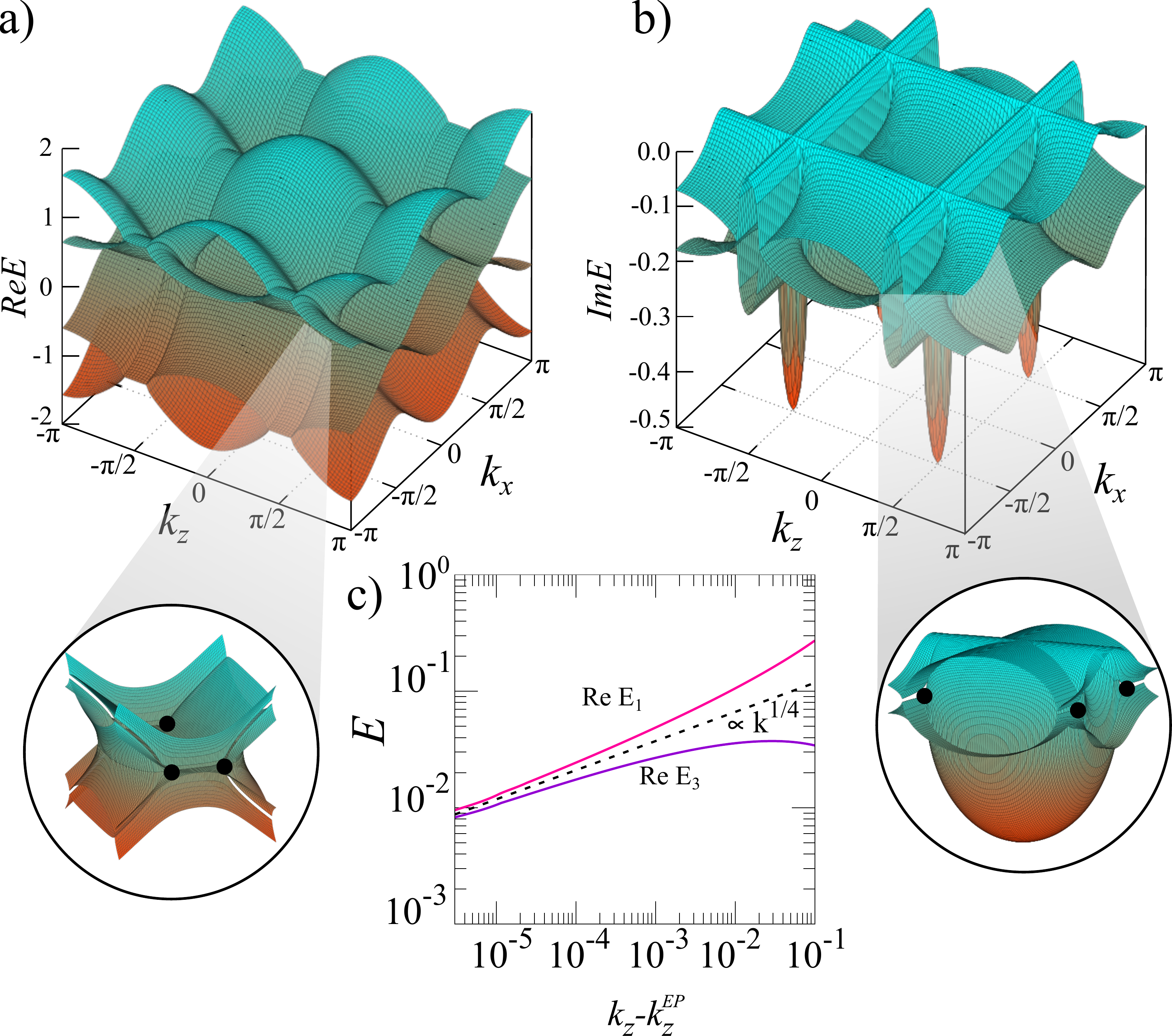}
  \caption{(color online) \textbf{a:} real and \textbf{b:} imaginary part of the energy eigenvalues at $k_{y}=\pi/2$. Four quadruplets of EP$^4$ located around [$k_{x},k_{z}$]\!=\![$\pm\pi/2,\pm\pi/2$] are visible; the band dispersion in the vicinity of one of those is represented in the insets. \textbf{c:} dispersion along the $k_{z}$ direction of the real part of the positive-energy eigenvalues close to one EP$^4$, compared with a quartic root behavior (dashed line).}
  \label{fig3}
\end{figure}

The DMFT self-energy gives rise to EP$^4$ in the band structure of the effective Hamiltonian, as illustrated in \figu{fig3}a and b, featuring respectively the real and imaginary part of the energy eigenvalues of $H_{\mathrm{eff}}(\mathbf{k})$ on the $k_{y}=\pi/2$ plane. In total, we have four quadruplets of EP$^4$ centered around $[k_{x},k_{z}]=[\pm \frac{\pi}{2},\pm\frac{\pi}{2}]$. The insets show the positions of the EP$^4$ around $\mathbf{k}=[\frac{\pi}{2},\frac{\pi}{2},\frac{\pi}{2}]$.  

\noindent
\textit{Hallmarks of the exceptional points} -- 
A $4^\text{th}$-root scaling behavior of the energy eigenvalues, close to the EP$^4$, is visible in \figu{fig3}c. The imaginary part (not plotted) displays an analogous behavior. The situation is identical in every $k$-direction and around each of the EP$^4$ of the model.
Even though it does not represents a necessary condition for their existence \cite{Demange2011}, this property clearly marks the presence of EP$^4$.

\begin{figure}[ht]
  \includegraphics[width=\linewidth]{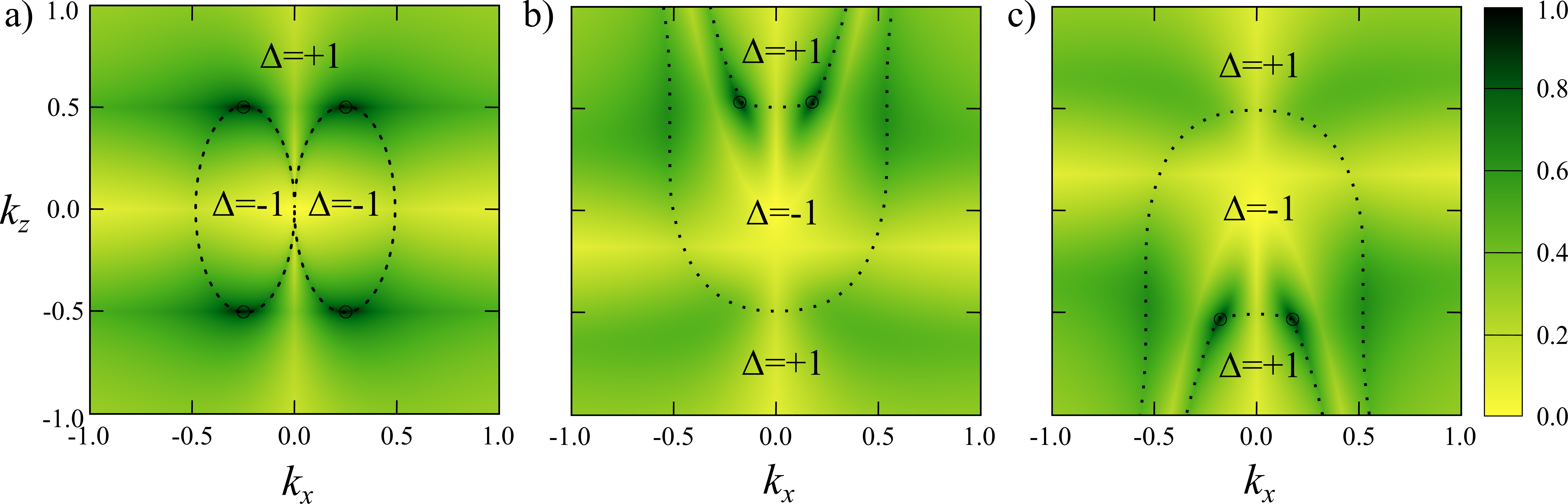}
  \caption{(color online) average value of the overlap of all possible couples of eigenvectors of Eq.~\eqref{nonhermitian_h} in the $k_{y}=0$ (a), $k_{y}=-\text{\textbabygamma}/4$ (b)  and 
  $k_{y}=+\text{\textbabygamma}/4$ (c) respectively. $\text{\textbabygamma}$  is set to $1/\sqrt{2}$. The black circles mark the position of the EP$^4$. The black dotted lines separate the different domains of the sign of the discriminant. }
  \label{fig4}
\end{figure}

Other considerations further strengthen the evidence for fourth-order exceptional points: as previously mentioned and as it can be seen in the right panel of \figu{fig1}, at the exceptional points, real and imaginary parts of all eigenvalues coincide and the Hamiltonian matrix becomes non-diagonalizable. A way to clearly visualize this is to plot the overlap between the eigenvectors of the effective Hamiltonian, that become degenerate at the EP$^4$. This is shown in \figu{fig4} for the linearized model \eqref{nonhermitian_h} on three different $k_{y}$ planes. The color scale indicates the average value of the overlap between all possible couples of eigenvectors: as expected, it takes precisely the value of 1 at the eight EP$^4$, denoted by black circles, where all eigenvectors coincide. 

A quantity that further characterizes the non-Hermitian effective system is $\Delta$, namely the sign of the discriminant of the characteristic polynomial $P(\text{-}iE)=\mathrm{det}(H(\mathbf{k})+iE)$. 
This has been suggested as an extension of the topological invariant $\mathrm{sgn}(\mathrm{det}H)$ in system with EP$^2$ ~\cite{yoshida2019,gong2018}.
The change of variable $E\rightarrow iE$ guarantees that $P$ has real coefficients only~\cite{delplace2021} so that a relation between the energy eigenvalues and the value of $\Delta$ can be established. 
As shown in \figu{fig4}, the sign of the discriminant delimits different regions in $k$-space. The borders (dotted contours) mark transition lines between different regimes in the eigenvalue spectrum.
To characterize them, consider an eigenvector of $H$, $\ket{v}$, with eigenvalue $E_{v}$; then its chiral-transformed is still an eigenvector with eigenvalue $-E^{*}_{v}$~\cite{Kawabata2019}. Hence, the two eigenvectors will coincide if the corresponding eigenvalues are purely imaginary. When $\Delta=-1$ (see for example the central region of \figu{fig1}b) $E_{1}$ and $E_{3}$ are indeed purely imaginary; the corresponding eigenmodes are then completely evanescent and invariant under chiral symmetry. On the contrary, when $\Delta=+1$, $E_{1}=-E_{2}^{*}$ and $E_{3}=-E_{4}^{*}$ all have nonzero real parts, and hence there are no chiral-invariant eigenvectors. Therefore, at the transition line, on which the EP$^4$ are bound to be located \footnote{See Supplemental Material for technical details.}, the chiral symmetry of the eigenvectors is spontaneously broken.

\noindent\textit{Conclusions} -- We have constructed a microscopic chiral symmetry preserving tight-binding model in three spatial dimensions that, in the presence of electronic interactions features fourth-order exceptional points. This has interesting observable implications regarding novel topological and analytical properties of the quasi-particle spectra found in correlated quantum materials.
While EP$^{4}$ have previously been observed in the context of waveguides, in this work we suggest the possibility of realizing them in chiral-symmetric electronic materials without any fine-tuning of parameters.
A possible route to realize the required -- purely orbital off-diagonal -- hopping structure in a solid is to consider ligand-mediated hopping between $d$-orbitals, as obtained from first principles in some iron-based superconducting materials \cite{Yin2011}. There, the orbital-diagonal nearest-neighbor hopping between transition-metal atoms arises from the superposition of the processes mediated by the pnictogen atom that, for specific geometries, can lead to perfectly destructive interference. 
In this situation, only the off-diagonal iron-iron hopping is finite, similar to what we suggest in our microscopic Hamiltonian.

\noindent\textit{Acknowledgments} -- We thank G. Rohringer for useful discussions. This work is funded by the Deutsche Forschungsgemeinschaft (DFG, German Research Foundation) through Project-ID 258499086-SFB 1170, Project-ID 247310070-SFB 1143 (Subproject A06), the W\"urzburg-Dresden Cluster of Excellence on Complexity and Topology in Quantum Matter –ct.qmat Project-ID 390858490-EXC 2147, and RE1469/13-1. 

\bibliography{references}

\end{document}